\documentclass[aps,prd,twocolumn, amsmath,superscriptaddress,amssymb,showpacs,floatfix,nofootinbib,longbibliography]{revtex4-1}
\pdfoutput=1
\usepackage{graphicx}
\usepackage{bm}
\usepackage{times}
\usepackage{slashed}
\usepackage{color}
\usepackage{aas_macros}
\usepackage{hyperref}
\usepackage{gensymb}
\hypersetup{
     colorlinks   = true,
     citecolor    = magenta,
     urlcolor     = magenta,
     linkcolor    = magenta
}

\usepackage{slashed}
\usepackage{lipsum}
\usepackage{subfigure}
\usepackage{multirow}
\usepackage{amsmath}
\usepackage{array} 
\usepackage{varwidth} 

\newcommand{\be}{\begin{equation}}
\newcommand{\ee}{\end{equation}}
\newcommand{\bea}{\begin{eqnarray}}
\newcommand{\eea}{\end{eqnarray}}

\begin{document}

\title{The time evolution of fast flavor crossings in post-merger disks around a black hole remnant}
\author{Payel Mukhopadhyay}
\email{pmukho@berkeley.edu; ORCID: orcid.org/0000-0002-3954-2005}
\affiliation{Departments of Astronomy and Theoretical Astrophysics Center, UC Berkeley, Berkeley, CA 94720, USA}

\author{Jonah Miller}
\email{jonahm@lanl.gov, ORCID: https://orcid.org/0000-0001-6432-7860}
\affiliation{CCS-2, Computational Physics and Methods, Los Alamos National Laboratory, Los Alamos NM 87544, USA}
\affiliation{Center for Theoretical Astrophysics, Los Alamos National Laboratory, Los Alamos NM 87544, USA}

\author{Gail C. McLaughlin}
\email{gcmclaug@ncsu.edu , ORCID: https://orcid.org/0000-0001-6811-6657}
\affiliation{Department of Physics, North Carolina State University, Raleigh, NC 27695 USA}

\begin{abstract}
 We postprocess a three-dimensional general
relativistic, full transport neutrino radiation magnetohydrodynamics simulation of the
black hole--accretion disk--wind system thought to be a potential outcome of the GW170817 merger to investigate the presence of electron lepton number (ELN-XLN) crossings in the neutrino angular distribution. Neutrinos are evolved with an explicit Monte Carlo method and can interact with matter via emission, absorption, or scattering. Within the postprocessing framework, we find ubiquitous occurrence of ELN-XLN crossings at early times ($\sim$ 11ms) but this does not hold for later times in the simulation. At postmerger times of $ \sim$ 60 ms and beyond, ELN-XLN crossings are only present near the equator. We provide a detailed analysis of the neutrino radiation field to investigate the origin and time evolution of these crossings. Previous reports have suggested ubiquitous flavor crossings persisting throughout the simulation lifetime, albeit for different sets of conditions for the merger remnant, the treatment of hydrodynamics and neutrino transport. Even though we do not perform a direct comparison with other published works, we qualitatively assess the reasons for the difference with our results. The geometric structure and evolution of the ELN-XLN crossings found in our analysis, and by extension, fast flavor instabilities have important implications for heavy element nucleosynthesis in neutron star mergers.  

\end{abstract}

\maketitle

\section{Introduction}

Following the kilonova detection in August 2017 (GW170817) \cite{LIGOScientific:2017vwq} which generated tremendous excitement in the scientific community, neutron star (NS) mergers became the first established site of the $r-$ process, a nucleosynthesis process which produces the heaviest elements in our universe \cite{RevModPhys.29.547,1976ApJ...210..549L,1977ApJ...213..225L,Metzger:2010sy,LIGOScientific:2017ync,Rosswog:2017sdn,1984SvAL...10..177B,Cote:2017evr}. As is thought to have happened for GW170817, a typical outcome of an NS merger is the formation of a central black hole or a hypermassive neutron star with a surrounding accretion disk \cite{Shibata:2017xdx,Sekiguchi:2011zd,Burns:2019byj}. Neutron-rich ejecta from these 
remnants
can
dominate the production of the $r-$ process elements in NS mergers \cite{Surman:2005kf,Radice_2018,PhysRevD.101.103002}. 

As with core-collapse supernovae (CCSNe), where neutrinos are the primary means by which gravitational energy is released enabling the stellar collapse \cite{janka_PhysicsCoreCollapseSupernovae_2016,muller_HydrodynamicsCorecollapseSupernovae_2020,mezzacappa_RealisticModelsCore_2022}, the transport of energy and lepton number by neutrinos play a key role in the evolution of the merger accretion disk \cite{radice_DynamicsBinaryNeutron_2020,sarin_EvolutionBinaryNeutron_2021}. Very importantly, for both of these systems, neutrino-matter interactions affect the composition of the ejecta by driving the evolution of the electron fraction ($Y_e$), a key parameter governing the resulting nucleosynthesis \cite{1994ApJ...433..229W,McLaughlin:1996yzo,kajino_CurrentStatusRProcess_2019}. Modeling neutrino emission, absorption and transport is therefore crucial for understanding the dynamical evolution and nucleosynthesis in these systems. 

Under the extreme conditions of CCSNe and NS mergers such as high densities and temperatures, neutrino self interactions can be high enough to cause rich, nonlinear flavor transformation phenomena, such as collective neutrino oscillations \cite{Duan:2005cp,duan_CollectiveNeutrinoOscillations_2010,chakraborty_CollectiveNeutrinoFlavor_2016,Patwardhan:2022mxg}, neutrino-matter resonance \cite{Malkus:2012ts,Malkus:2014iqa,vaananen_UncoveringMatterneutrinoResonance_2016}, fast flavor instabilities (FFI) \cite{sawyer_SpeedupNeutrinoTransformations_2005,sawyer_MultiangleInstabilityDense_2009,sawyer_NeutrinoCloudInstabilities_2016,chakraborty_SelfinducedFlavorConversion_2016,izaguirre_FastPairwiseConversion_2017,chakraborty_SelfinducedNeutrinoFlavor_2016,tamborra_FlavordependentNeutrinoAngular_2017,dasgupta_FastNeutrinoFlavor_2017,dasgupta_SimpleMethodDiagnosing_2018,dasgupta_FastNeutrinoFlavor_2018,abbar_FastNeutrinoFlavor_2018,Abbar:2018shq,Yi:2019hrp,Martin:2019gxb,abbar_FastNeutrinoFlavor_2019,Martin:2021xyl,Johns:2020qsk,johns_FastFlavorInstabilities_2021,nagakura_NewMethodDetecting_2021,nagakura_WhereWhenWhy_2021,harada_ProspectsFastFlavor_2022,shalgar_NeutrinoPropagationHinders_2020,shalgar_OccurrenceCrossingsAngular_2019,shalgar_DispellingMythDense_2021,padilla-gay_FastFlavorConversion_2021,padilla-gay_MultiDimensionalSolutionFast_2021,shalgar_ThreeFlavorRevolution_2021,hansen_EnhancementDampingFast_2022} and collisional instabilities \cite{johns_CollisionalFlavorInstabilities_2021,Johns:2022yqy}.  Of these the fast flavor instabilities tend to have faster growth rates and tend to occur closer to the center than collective oscillations and the neutrino matter resonance. Ref. \cite{capozzi_FastFlavorConversions_2017} studied FFI in a two beam model, and based on these results, speculated that FFI might occur in supernovae. Ref. \cite{nagakura_FastpairwiseCollectiveNeutrino_2019,harada_ProspectsFastFlavor_2022} showed regions of  FFI that occur inside the region of the shock in supernovae models.  This suggests that FFI could potentially change the dynamics of the explosion. FFI have been reported to occur \textit{ubiquitously} in
NS mergers \cite{wu_FastNeutrinoConversions_2017, wu_ImprintsNeutrinopairFlavor_2017a,li_NeutrinoFastFlavor_2021,george_FastNeutrinoFlavor_2020,richers_EvaluatingApproximateFlavor_2022,just_FastNeutrinoConversion_2022} at extremely fast (nanosecond) timescales. It has also been shown that FFI can have significant impact on the composition of the ejecta and the corresponding nucleosynthesis. In \cite{li_NeutrinoFastFlavor_2021}, it was found that ubiquitous FF oscillations result in a significantly more neutron-rich outflow, providing lanthanide and 3rd-peak r-process abundances similar to solar abundances. 

Recognizing where FFI will take place in NS mergers is therefore very valuable, however, given the fast timescales, and short length scales, of FFI, global simulations of NS mergers incorporating FFI from first principles is computationally intractable. However, extensive studies have shown that existence of FFI necessitates that at a given location, there needs to be an overabundance of neutrinos moving in some directions, while an overabundance of antineutrinos moving in other directions \cite{dasgupta_FastNeutrinoFlavor_2017,abbar_FastNeutrinoFlavor_2018,morinaga_FastNeutrinoFlavor_2022}. Morinaga et al. has presented an argument that a \textit{necessary and sufficient} condition for FFI is that the angular distributions of neutrinos and antineutrinos must cross each other at a given point in space \cite{morinaga_FastNeutrinoFlavor_2022}. This criteria is known as the existence of an electron lepton number (ELN-XLN) crossing. Classically evolved neutrino fields can be tested in postprocessing for the presence of such crossings, e.g. \cite{richers_EvaluatingApproximateFlavor_2022}, as well as for the growth rate of the instability using linear stability analysis \cite{Froustey:2023skf}.  While such postprocessing tests do not capture the true nature of the neutrino flavor field because they do not include the effect of the flavor transformation that occurs simultaneously with the advection of the neutrinos, they are a useful first step in determining where to focus efforts on the neutrino quantum kinetics.

In this paper, we postprocess the general relativistic, neutrino radiation magnetohydrodynamics (GR$\nu$RMHD) simulation of a black-hole accretion disk-wind system performed in \citep{miller_FullTransportModel_2019} using the $\nu\texttt{bhlight}$ code \citep{nubhlight, miller_FullTransportModel_2019, miller_FullTransportGeneral_2020}. This model was originally selected to match a potential outcome of the GW170817 event. Evolved using the Monte Carlo method, we analyze the full neutrino distribution and examine where ELN-XLN crossings form. We analyze different times in the simulation and present our results taking an early ($\sim 11$ ms) and a later ($\sim 65$ ms) time snapshot as representative examples. 

Analysis of the conditions conducive to FFI in compact object merger remnants have been reported previously. Early estimates used classical $\nu_e$ and $\bar{\nu}_e$ emission surfaces to be two separate rectangular prisms \cite{wu_FastNeutrinoConversions_2017} or tori  \cite{wu_ImprintsNeutrinopairFlavor_2017a} together with a dispersion relation approach based on linear stability analysis to find that conditions favorable to FFI were present almost everywhere above the neutrinosurface. Ref. \cite{just_FastNeutrinoConversion_2022} investigated the occurrence of ELN-XLN crossings within the postprocessing approximation for a black hole accretion disk where neutrino transport was based on angular moments of the neutrino distribution using the ELN-XLN finding algorithm developed in \cite{abbar_SearchingFastNeutrino_2020} for moment-based approaches, finding crossings nearly everywhere above the neutrino surface. Postprocessing approaches have also been carried out to detect ELN-XLN crossings in the context of hypermassive neutron star accretion disks \cite{richers_EvaluatingApproximateFlavor_2022} for Monte Carlo based methods where more detailed information of the neutrino angular distribution is available. These studies also suggest the presence of ubiquitous ELN-XLN crossings.  An angular moment-based technique for stability analysis was developed by \cite{Froustey:2023skf} and applied to a snapshot of a neutron star merger simulation and compared with the spatial extent of ELN-XLN crossings.  They also found that most neutrinos would encounter a fast flavor instability on their way out of the merger.

Ref. \cite{li_NeutrinoFastFlavor_2021} was the first simulation to incorporate FFI transformation dynamically into a simulation with a black hole accretion disk. They also found ELN-XLN crossings to be practically present across the entire simulation zone. This method employs an underlying classical neutrino transport calculation, tests for the presence of an instability and then redistributes the classical neutrinos fields using an ad-hoc description.  The results suggested that the FFI could have significant impact on the resulting element synthesis.  Ref. \cite{just_FastNeutrinoConversion_2022} also implemented FFI conversion dynamically in a black hole  accretion disk scenario and also found an increase in r-process nucleosynthesis yields. In a related work Ref. \cite{Fernandez:2022yyv} looked at outflows from a hypermassive neutron star remnant in the presence of FFI and again found increases in the amount of r-process element synthesis.

Many questions remain open regarding FFI.  Most obviously, we do not yet have a fully dynamical evolution of a neutron star merger in the presence of neutrino flavor transformation.  However, whether all neutrinos produced in a remnant will undergo an FFI has not been definitively determined.  In this work, we take a step in this direction.  For the first time, we perform a detailed analysis of the neutrino radiation field to examine the physical processes giving rise to ELN-XLN crossings in the context of Monte Carlo radiation transport around a black hole accretion disk. We find that the spatial regions where ELN-XLN crossings exist can evolve: they are ubiquitous at early times ($\sim 11$ ms), but only occur near the equator at later ($\sim 65$) times. We note at the outset that while our findings are limited by the postprocessing approach as compared treating oscillations self-consistently \cite{li_NeutrinoFastFlavor_2021}, it also has the significant advantage of having access to the full neutrino distribution as a consequence of the Monte Carlo method and the radiation field being generated \textit{self-consistently, inline with the simulation}. Our study should motivate further investigations into uncovering the nature of ELN-XLN crossings and its implications for the nature of FFI and the corresponding $r-$ process nucleosynthesis in NS mergers.

\section{Model}

\begin{figure*}
    \centering
    \includegraphics[width=\columnwidth]{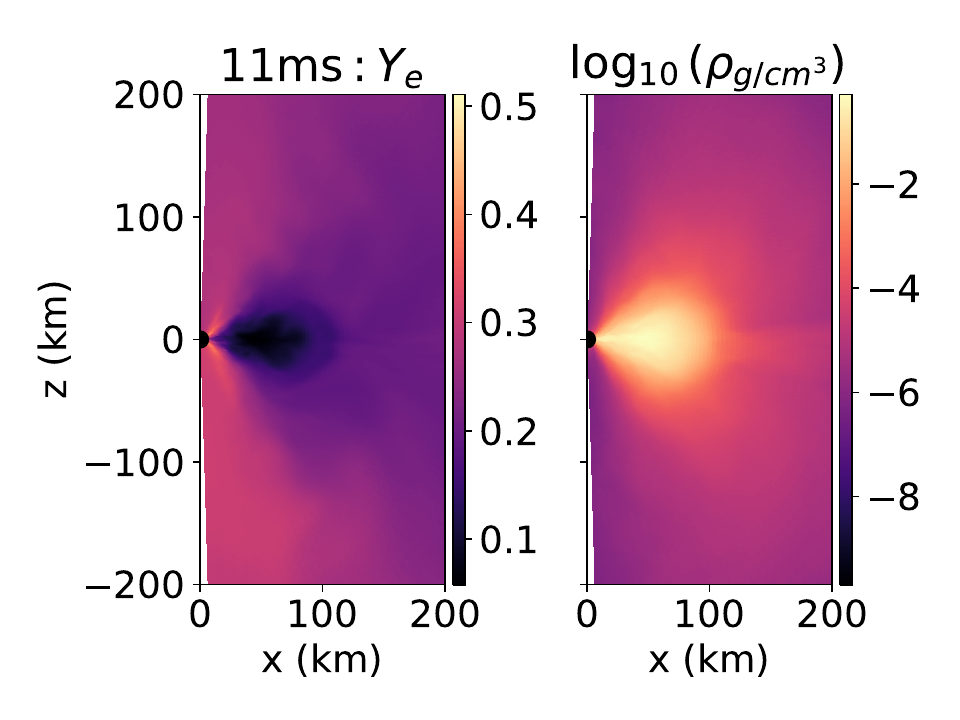} \\
    \includegraphics[width=\columnwidth]{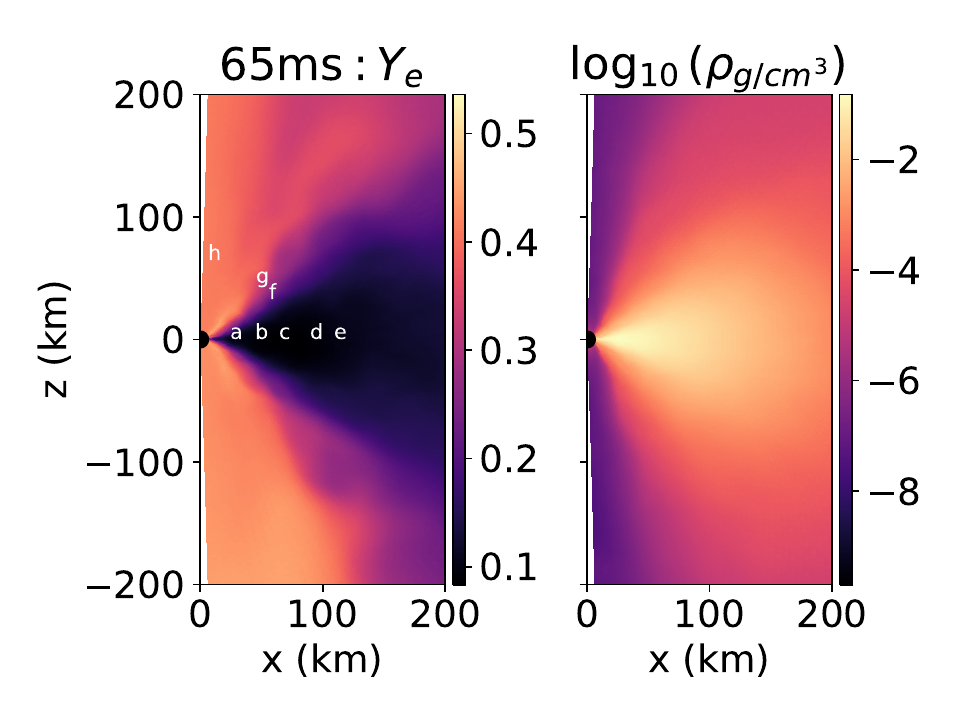}
    
    \caption{Density and electron fraction of the simulation at $t \sim$ 11 ms and $t \sim$ 65 ms. The text labels in the 65 ms $Y_e$ plot are the locations where the representative crossings analysis are shown in Figs. \ref{fig:65msequator} and \ref{fig:65msangular}. }
    
    \label{fig:densityandye}
\end{figure*}

This model uses a stationary Kerr \citep{KerrBH} black hole
spacetime for a black hole of mass $M_{\rm BH} = 2.58 M_\odot$
and dimensionless spin $a=0.69$. The initial conditions are
a torus in hydrostatic equilibrium \citep{FishboneMoncrief} of
constant specific angular momentum, constant entropy of
$s = 4 k_b/$baryon, constant electron fraction $Y_{\rm e}=0.1$, and total mass of $M_{\rm d} = 0.12 M_\odot$. The angular momentum is set by the radius of maximum pressure, which is 10.46 $G M_{\rm BH}/c^2$. The mass is set by the inner radius of the torus, which is 5 $G M_{\rm BH}/c^2$. The torus starts with a single poloidal magnetic field loop. The initial magnetic field strength scales with disk density and is normalized such that the ratio of gas to magnetic pressure, $\beta$, is 100 at the radius of maximum pressure, corresponding to roughly $1.7\times 10^{14}$ Gauss. As the disk evolves, the magnetorotational instability \citep{Velikov1959,Balbus1991} amplifies the magnetic field. At saturation, the peak field strength is roughly $3\times 10^{15}$ Gauss. We note that with sufficient time, this saturation value will always be achieved, even if a smaller initial field strength is chosen.

$\nu\texttt{bhlight}$ solves the equations of general relativistic ideal magnetohydrodynamics, closed with the SFHo equation of state, described in \citep{SFHoEOS} and tabulated in \citep{stellarcollapsetables}.
Neutrinos are evolved with an explicit Monte Carlo method and can interact with matter via emission, absorption, or scattering. As with most state-of-the-art models, $\mu$ and $\tau$ neutrinos and their antiparticles are grouped together as so-called \textit{heavy} neutrinos. The number density of heavy neutrinos is about two orders of magnitude lower than the number density of electron neutrinos or their antiparticles. 
For emission and absorption, interactions are tabulated in \citep{fornax} and summarized in \citep{BurrowsNeutrinos}. Neutrino scattering is implemented as described in \citep{nubhlight}. The Monte Carlo and Finite Volume methods are coupled via first-order operator splitting.

The simulation was performed on a radially logarithmic, quasi-spherical grid in horizon penetrating coordinates with $N_r\times N_\theta \times N_\phi = 192\times 168\times 66$ grid points with approximately $3.8\times 10^7$ Monte Carlo packets. The simulation was run for a physical time of 10,000 $G M_{BH}/c^3$, or approximately 130ms. Fig. \ref{fig:densityandye} shows $Y_e$ and $\rho$ for two representative time snapshots. As can be seen from the figure, for both 11 ms and 65 ms, the outer parts of the disk are protonizing due to $\bar{\nu_e}$ emission while the inner high-density regions of the disk stay strongly neutron rich. For more details on the original simulation and analysis thereof, see \citet{miller_FullTransportGeneral_2020}.
\begin{figure*}[htb!]
    \centering
\includegraphics[width=\columnwidth]{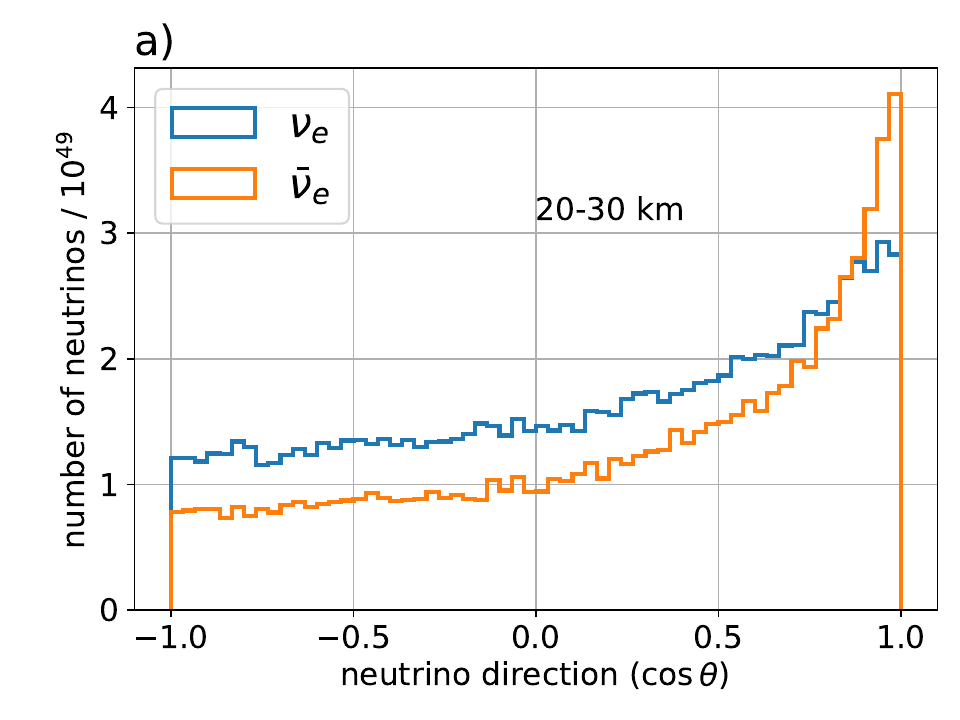}
   \includegraphics[width=\columnwidth]{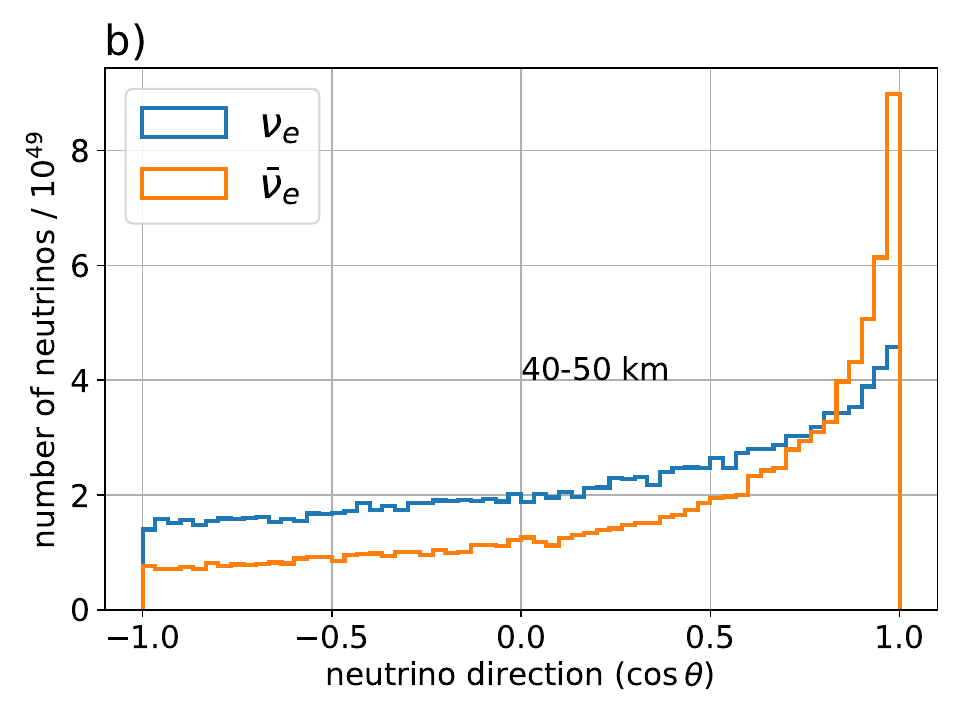}
    \includegraphics[width=\columnwidth]{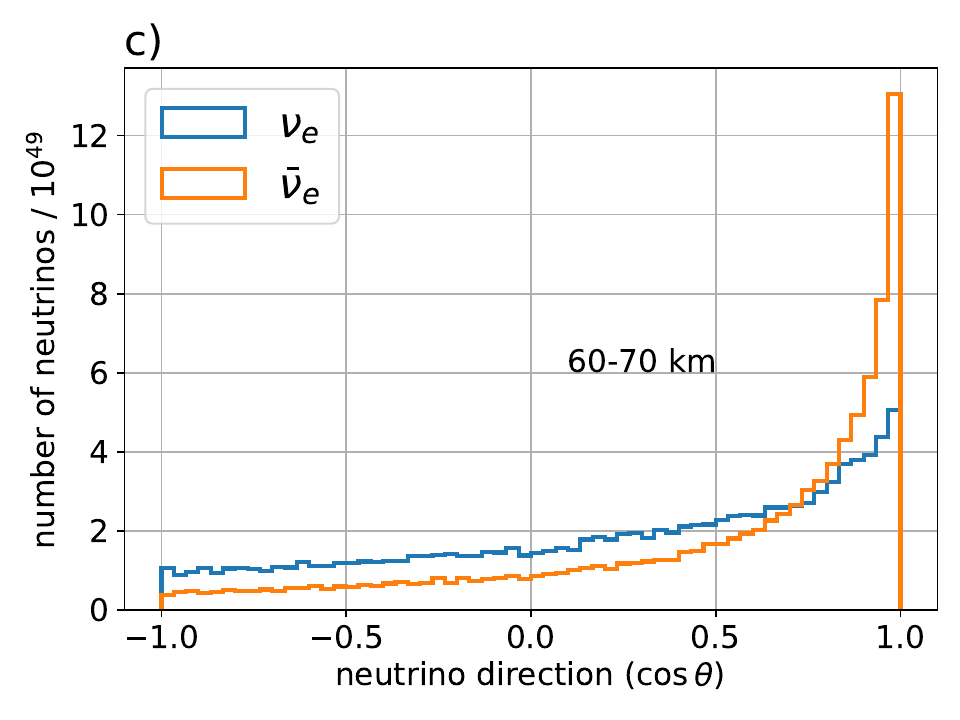}
    \includegraphics[width=\columnwidth]{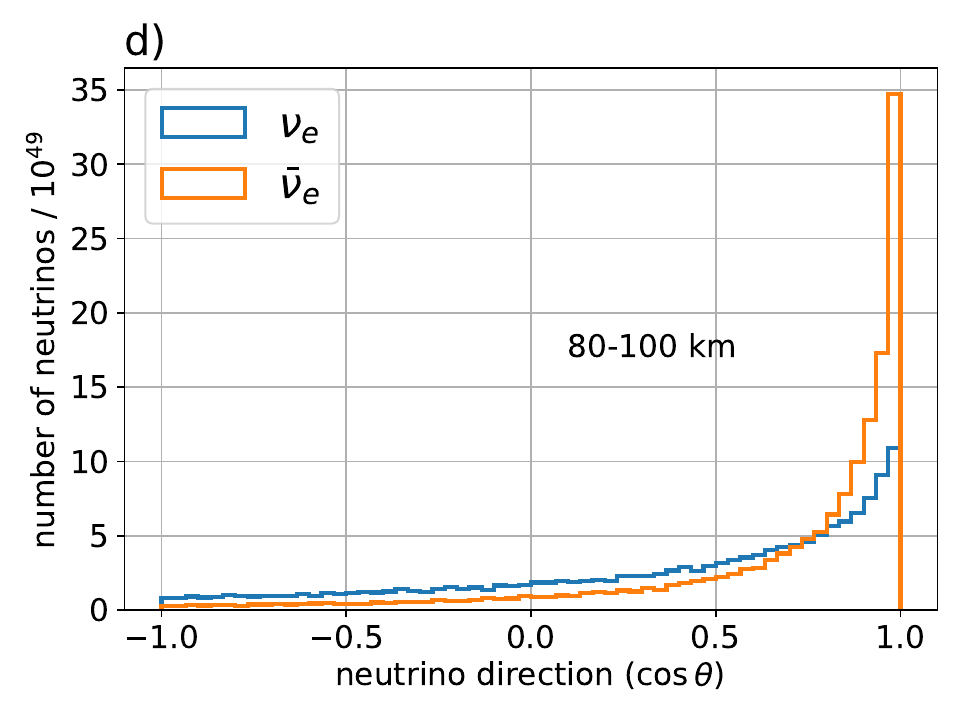}
    \includegraphics[width=\columnwidth]{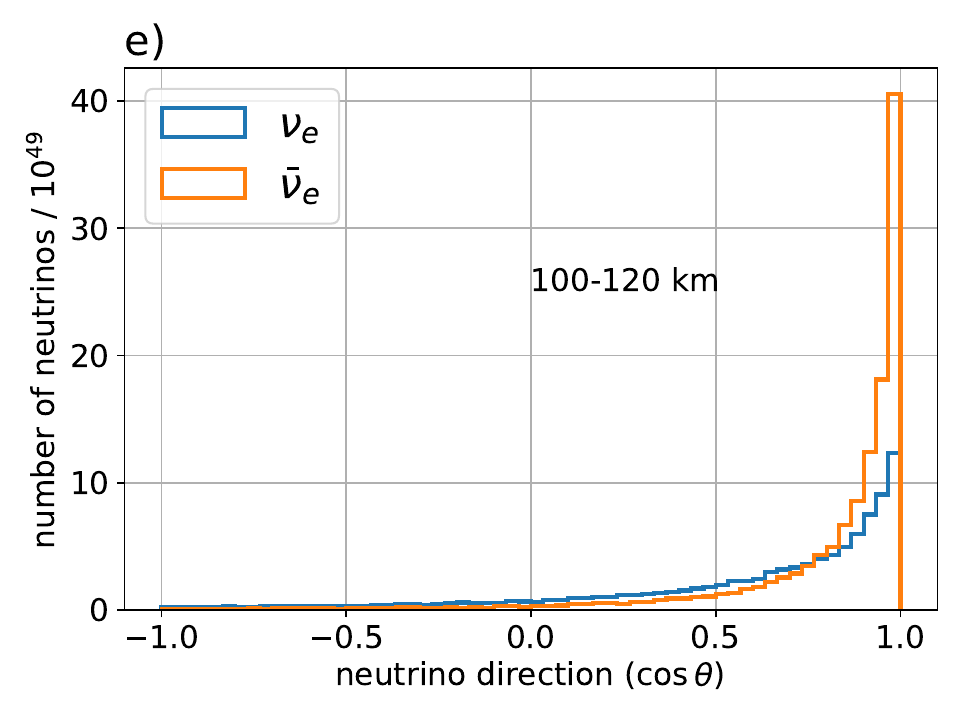}
    
    \caption{Angular distribution of $\nu_e$ and $\bar{\nu}_e$ across the disk at $t \sim 65$ ms. $\cos{\theta}$ is defined with respect to the radial direction so that $\cos{\theta}=1$ are neutrinos moving radially outwards. We find robust crossings between the $\nu_e$ and $\bar{\nu}_e$ distributions across the entire disk between $r \sim$ 20--120 km beyond which the radiation field turns off. The titles of the plot a), b) c), d) and e) correspond to the respective spatial locations labeled in Fig. \ref{fig:densityandye}.}
    \label{fig:65msequator}
\end{figure*}

We analyze the neutrino radiation field, which was generated self-consistently, inline with the simulation, but without oscillations and look for evidence of the ELN-XLN crossings. To build sufficient statistics for the neutrino distribution function from the Monte Carlo data, we consider control volumes ($dV$) in the physical space that typically contain some small range in radius $r$ and polar angle $\theta^{polar}$ but include the entire azimuthal angle $\phi$. This enables us to build up excellent statistics on our Monte Carlo particles. To make contact to the fluid variables in a consistent way, we azimuthally average our fluid grid data to provide a representative fluid field at a given $r$ and $\theta^{polar}$.

\section{Results}

In this section, we report on the geometric structure of ELN-XLN crossings and discuss how and why they evolve with time.

We first examine the angular distribution of electron neutrinos and electron antineutrinos at 65 ms at the various spatial locations that are indicated in the snapshot in Fig. \ref{fig:densityandye}.  In Fig.  \ref{fig:65msequator}
each subplot corresponds to a different radial distance from the black hole along the equator, corresponding to points a through e in Fig. \ref{fig:densityandye}. Fig. \ref{fig:65msequator} shows the number distributions of $\nu_e$ and $\bar{\nu}_e$'s 
as a function of the angle $\theta$ between the neutrino velocity vector and the radial direction such that $\cos{\theta}=1$ corresponds to neutrinos moving radially outwards. The neutrino number distributions inside $dV$ are extracted by analyzing the Monte Carlo particles in the simulation data. Each plot in Fig. \ref{fig:65msequator} corresponds to a volume element located along the equator of the disk with the radial range ($dr$) of the volume element indicated on the plot.  The $\nu_e$ and $\bar{\nu}_e$ number distributions summed over this volume element are shown. ELN-XLN crossings can be robustly seen across the entire radial range near the equatorial region of the disk, suggesting that the disk can host FFI in NS mergers. 

\begin{figure*}
    \centering
\includegraphics[width=\columnwidth]{60-70km_fast_flavor_equator_radial}
   \includegraphics[width=\columnwidth]{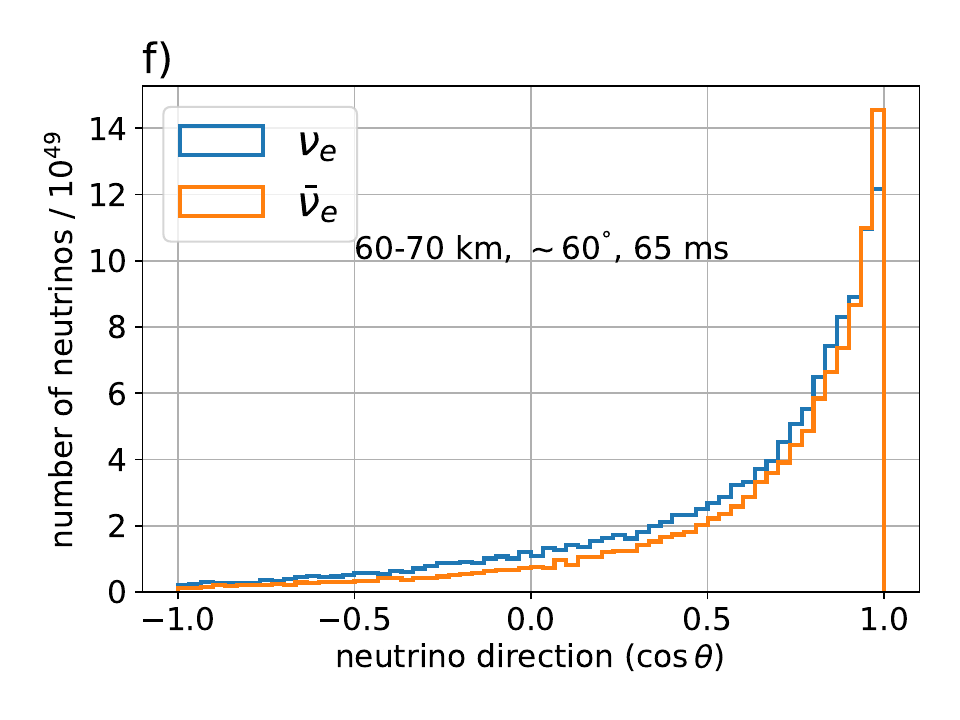}
    \includegraphics[width=\columnwidth]{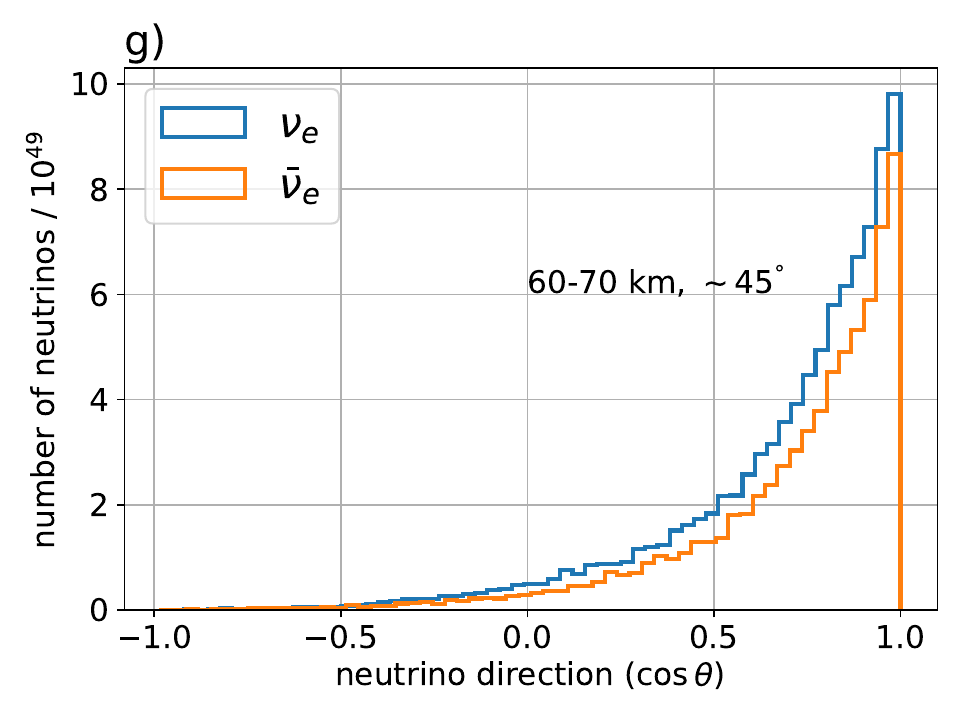}
    \includegraphics[width=\columnwidth]{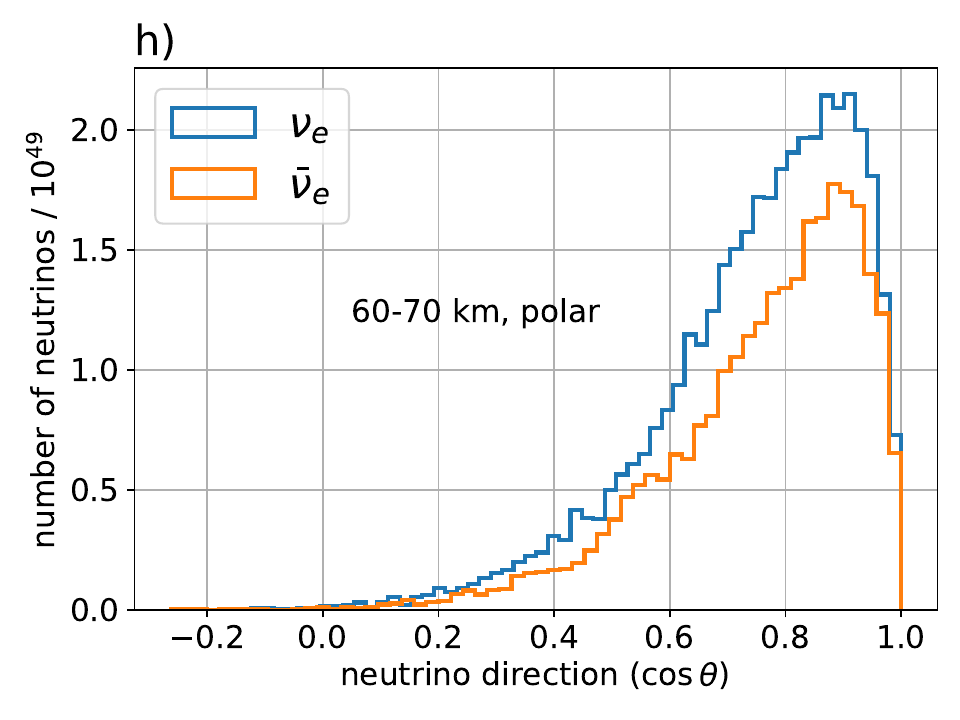}
    
    \caption{Evolution of the  crossings as a function of angle from the pole. The discretized radial distance of a neutrino packet is set to be 60--70 km. The top two plots are at  (equator) and $60\degree$ from the pole (30$\degree$ from equator). Clear crossings are present at and $30\degree$ above the equator. As one moves further above the disk as shown in the bottom panels, the crossings disappear and $\nu_e$ distribution overwhelms $\bar{\nu}_e$'s. Crossings have already disappeared by the time the neutrinos are $45 \degree$ from the equator. The titles of the plot c), f), g) and h) correspond to the respective spatial locations labeled in Fig. \ref{fig:densityandye} as one moves up from the equator.}
    \label{fig:65msangular}
\end{figure*}

In Fig. \ref{fig:65msangular}, the neutrino distributions are plotted as one moves away from the disk towards the pole.  The plots in Fig. \ref{fig:65msangular} correspond to the locations c, f, g and h in Fig. \ref{fig:densityandye} where the radial size of the volume packet $dV$ is fixed between 60--70 km ($dr = 10~\mathrm{km}$) and the angular dependence ($\theta^{\rm polar}$) is explored. The top two panels show clear crossings at the equator and 30$\degree$ above the equator, but the crossings disappear as one moves further away from the equator. As shown in the bottom left plot, the crossing has disappeared by 45$\degree$ as $\nu_e$s overwhelm $\bar{\nu}_e$s at small $\theta^{\rm polar}$.

The right panel of Fig. \ref{fig:crossingoverview} summarizes the results so far, ELN-XLN crossings are only present around a $\sim$ 30$\degree$ angular range around the disk at $t \sim 65 \mathrm{ms}$. This plot was obtained by searching for crossings across the whole simulation volume. The volume was divided with radial resolution of $\sim 10$ km and angular resolution of about $6\degree$. The calculations were run on a high performance computing cluster. Green shaded regions are where crossings are present while white regions are the no-crossing zones.

\begin{figure*}
    \centering
\includegraphics[width=\columnwidth]{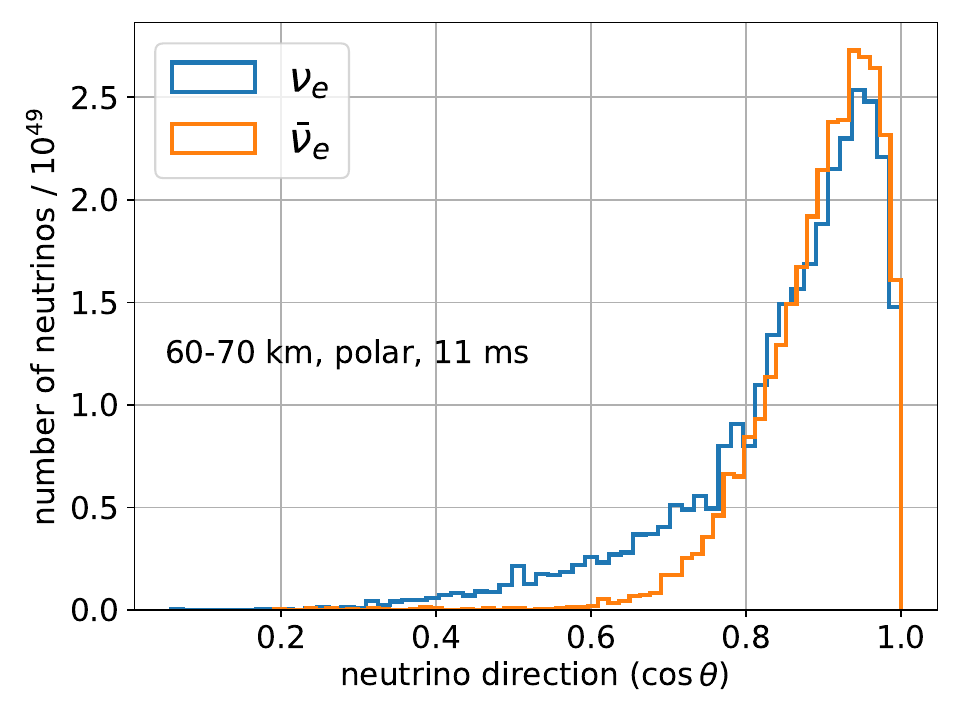}
   \includegraphics[width=\columnwidth]{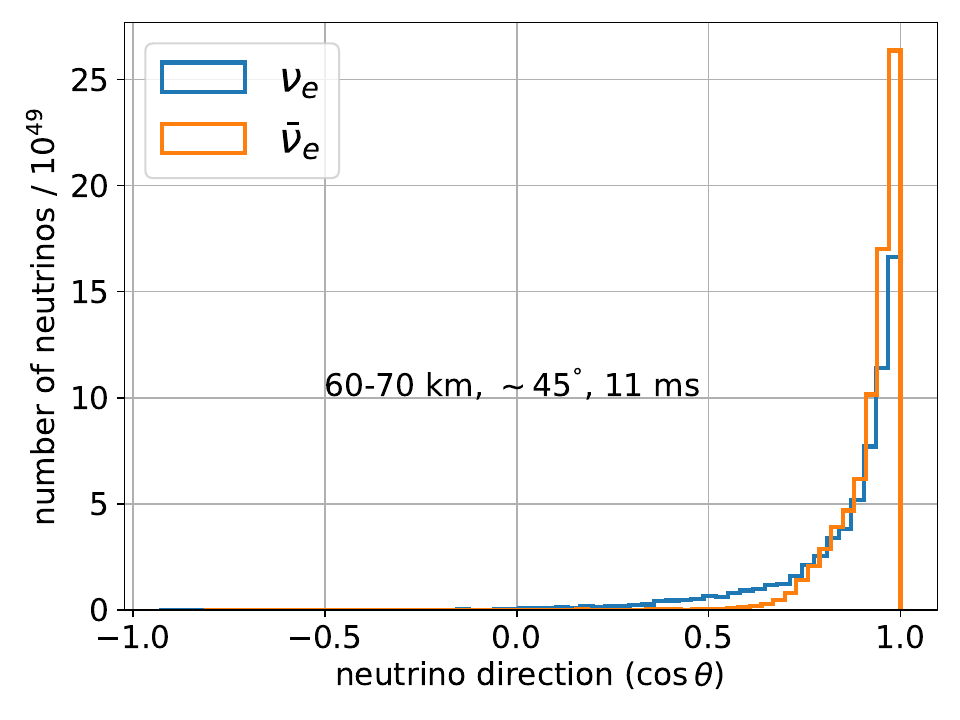}
    \caption{ Crossings at $45 \degree$ and polar regions for the 11ms snapshot. Compared to the results at $t\sim$65 ms where the crossings only occur close to the disk, they occur ubiquitously for earlier times.}
    \label{fig:11msangular}
\end{figure*}

\begin{figure*}[t]
    \centering
    \includegraphics[width=\columnwidth]{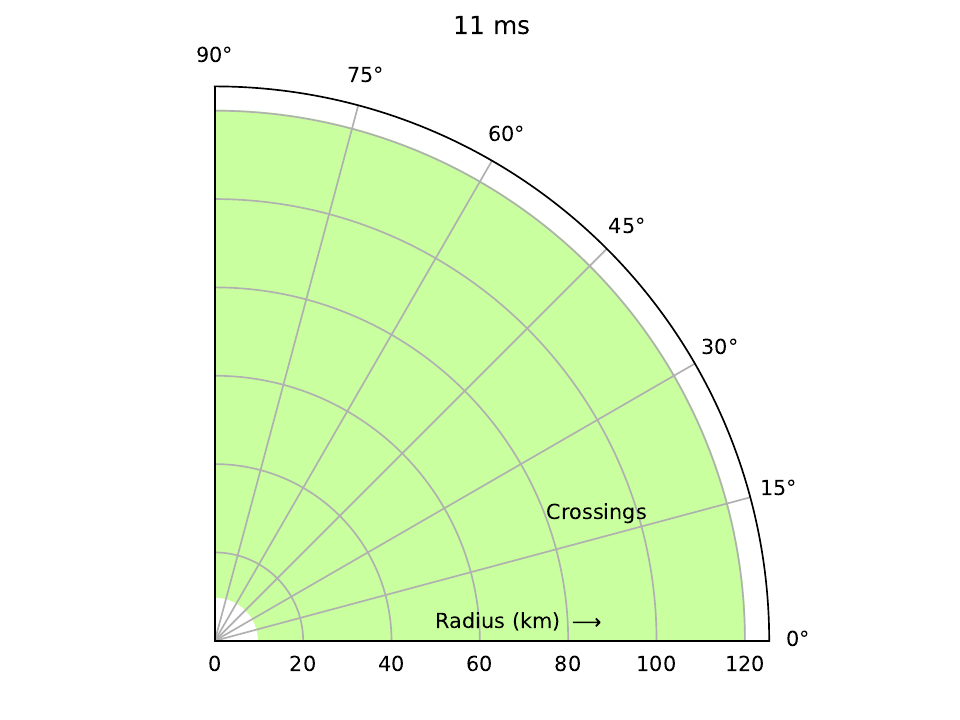}
    \includegraphics[width=\columnwidth]{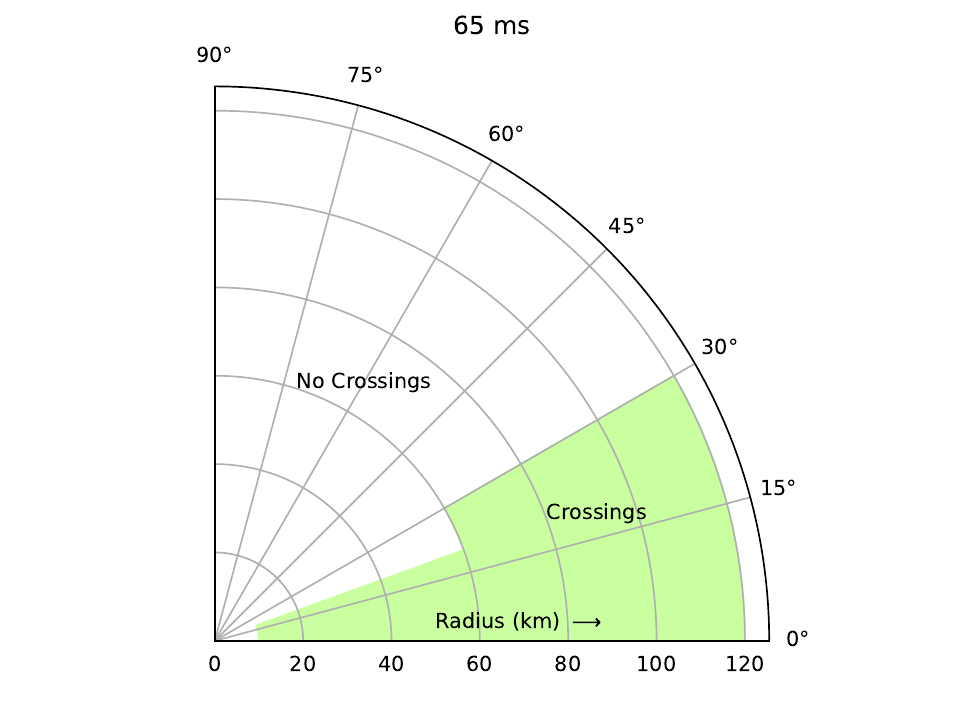}
    
    \caption{Locations in the $r-\theta$ plane where regions hosting crossings crossings are shaded in green. As explained in the text, crossings are ubiquitous at early times ($\sim 11$ ms), but then evolve to only occur closer to the disk as time evolves ($\sim 65$ ms).}
    
    \label{fig:crossingoverview}
\end{figure*}

Before delving into this geometric structure, we first explore the time dependence of these crossings by performing the same analysis on a $t\sim$ 11 ms snapshot as shown in Fig. \ref{fig:11msangular}. Compared to the bottom two plots of Fig. \ref{fig:65msangular}, clear crossings in the 45$\degree$ and polar region can be seen at 11 ms.  This demonstrates that the geometric structure of the ELN-XLN crossings shown in Figs. \ref{fig:65msequator} and \ref{fig:65msangular} is time dependent.  In fact, in this simulation at 11 ms ELN-XLN crossings are ubiquitous, as can be seen in the right panel of Fig. \ref{fig:crossingoverview}.

\begin{figure*}[htb!]
    \centering

     \includegraphics[width=\columnwidth]{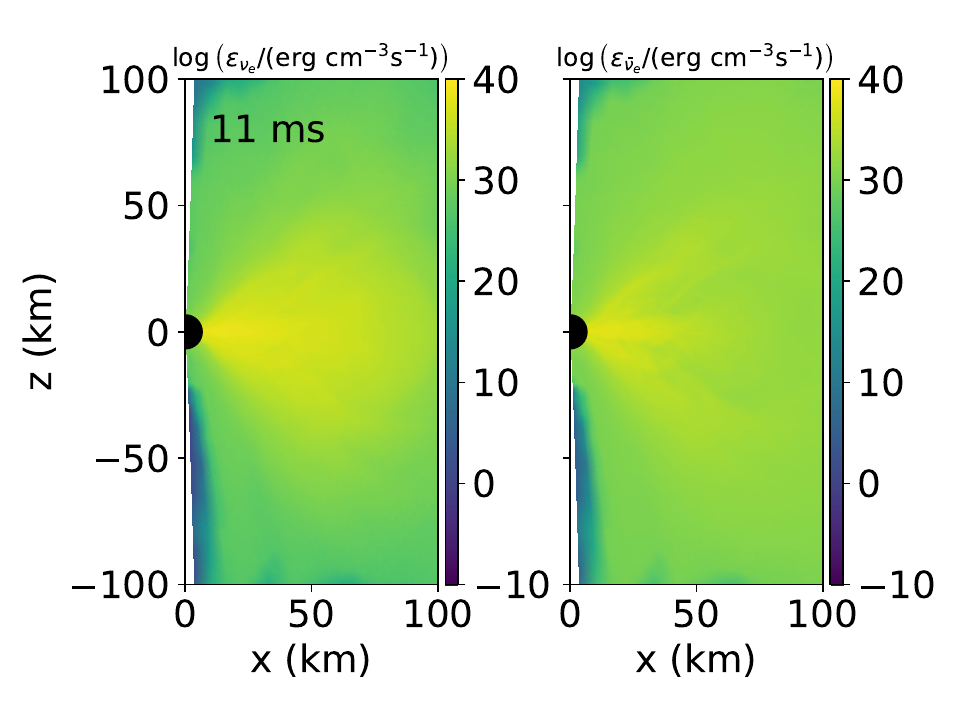}
   \includegraphics[width=\columnwidth]{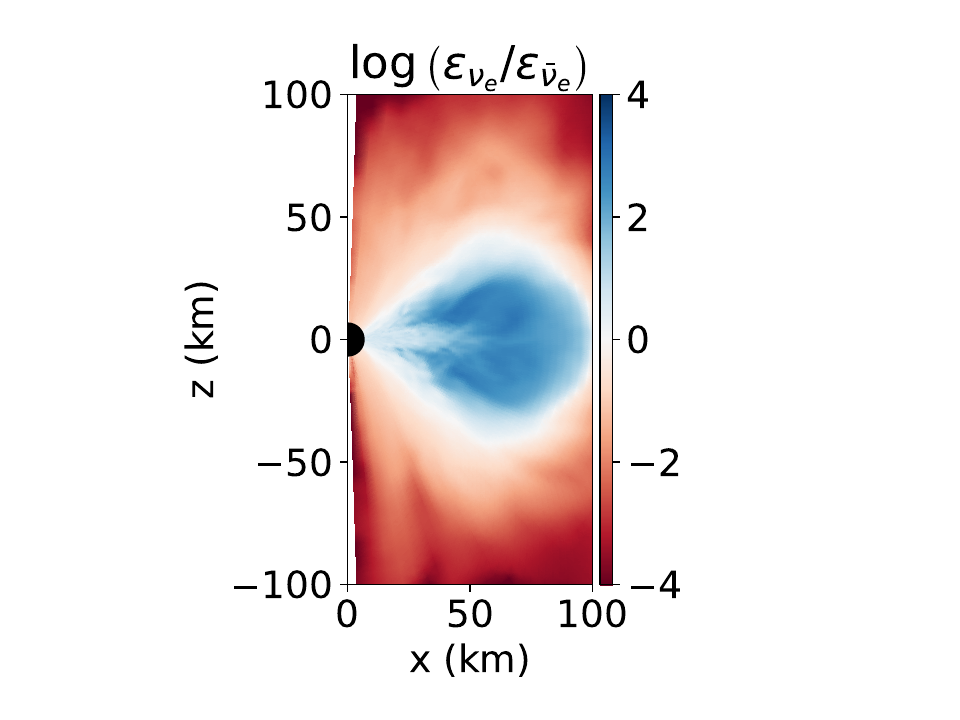}
   
\includegraphics[width=\columnwidth]{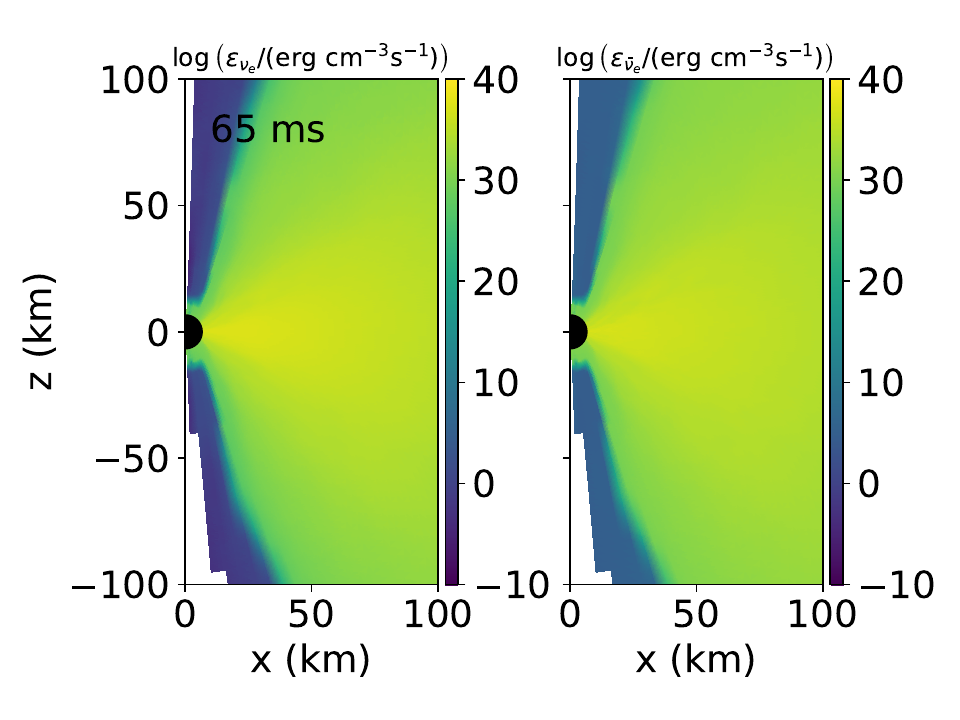}
   \includegraphics[width=\columnwidth]{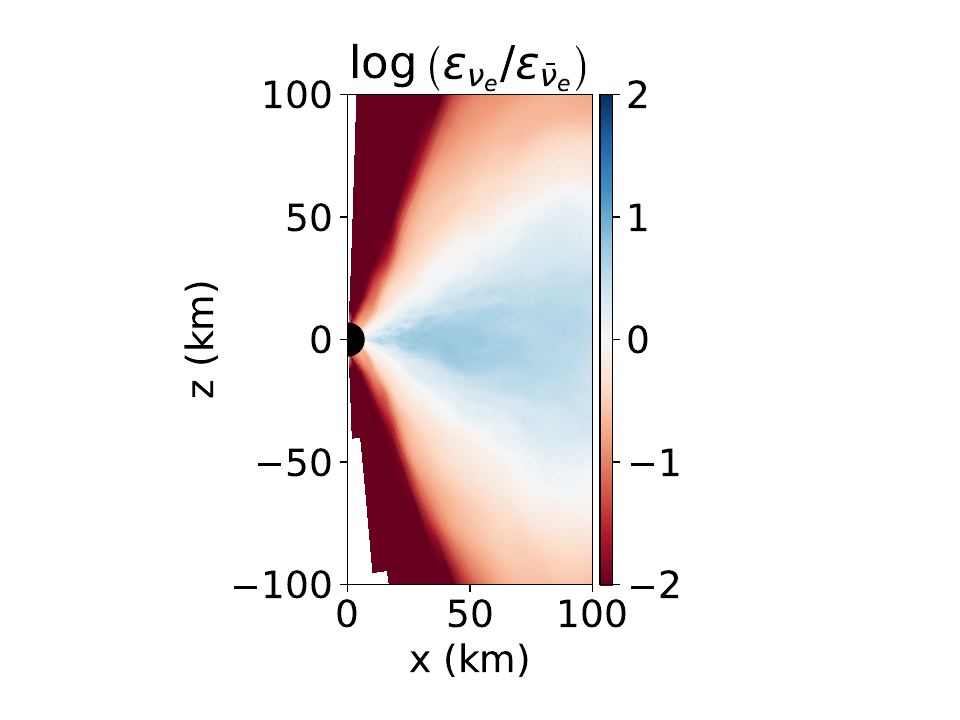}
      
    \caption{Top panel: 11ms snapshot. First two panels show individual emissivities ($\epsilon$) of $\nu_e$ and $\bar{\nu}_e$. Absolute values of the emissivities are highest near the disk. The rightmost panel shows $\frac{\epsilon_{\nu_e}}{\epsilon_{\bar{\nu}_e}}$. At these early times, $\epsilon_{\nu_e}$ can be up to $\sim$ 4 orders of magnitude greater than $\epsilon_{\bar{\nu}_e}$ close to the disk. Geometry wise, $\epsilon_{\nu_e}$ has a different structure than $\epsilon_{\bar{\nu}_e}$. As one moves away from the disk towards the poles, the individual emissivities drop, and $\epsilon_{\bar{\nu}_e}$ dominates over $\epsilon_{\nu_e}$. Bottom panel: 65 ms snapshot. The emissivities decrease with time for both $\nu_e$ and $\bar{\nu}_e$. The bottom right panel also shows that the emission structure has changed as compared to 11 ms case. Compared to the 11 ms case, the geometric emission structure of $\nu_e$ and $\bar{\nu}_e$ are more similar. $\epsilon_{\nu_e}$ still dominates over $\epsilon_{\bar{\nu}_e}$ close to the disk, but now the ratios differ by at most one order of magnitude near the disk. As explained in the text, this emission structure for 11 ms and 65 ms are crucial for determining the presence of crossings.}
    \label{fig6}
\end{figure*}

\begin{figure*}
    \centering
     \includegraphics[width=\columnwidth]{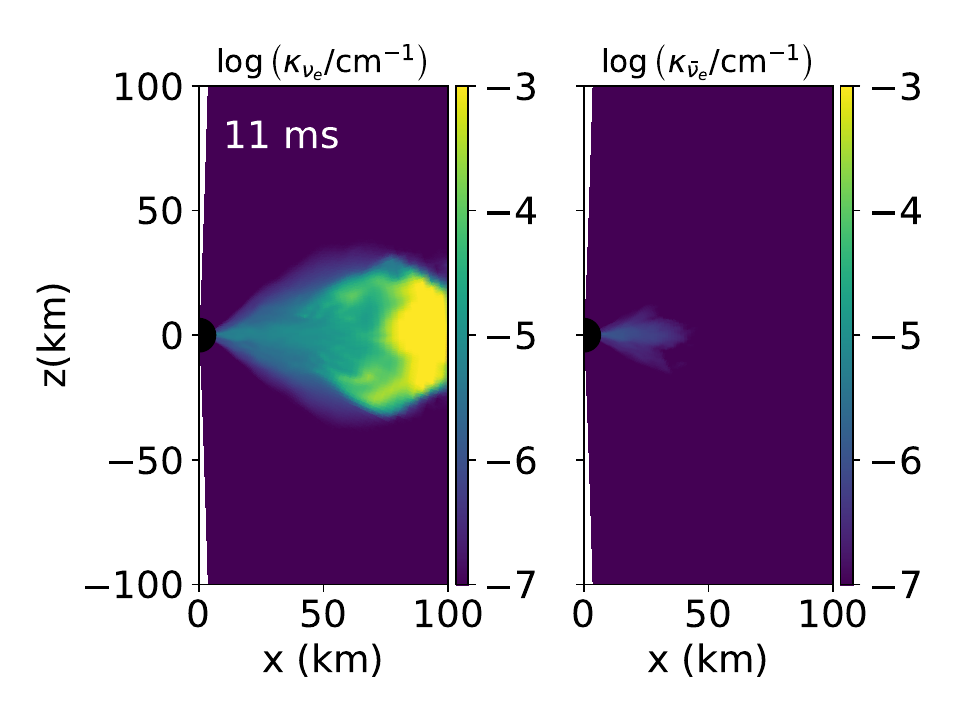}
     \includegraphics[width=\columnwidth]{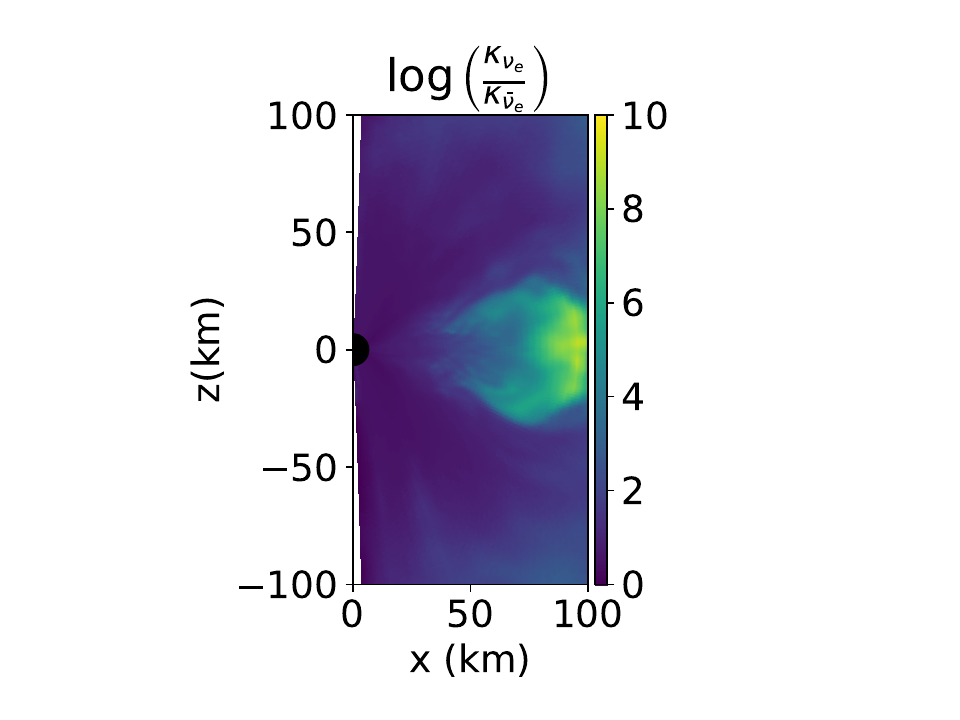}
\includegraphics[width=\columnwidth]{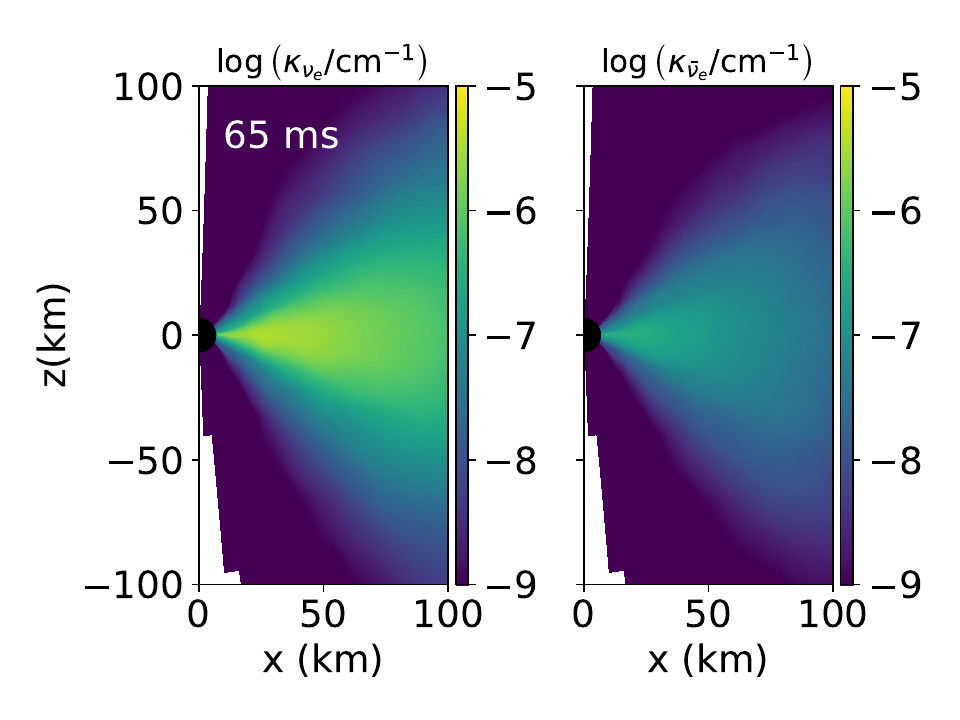} 
\includegraphics[width=\columnwidth]{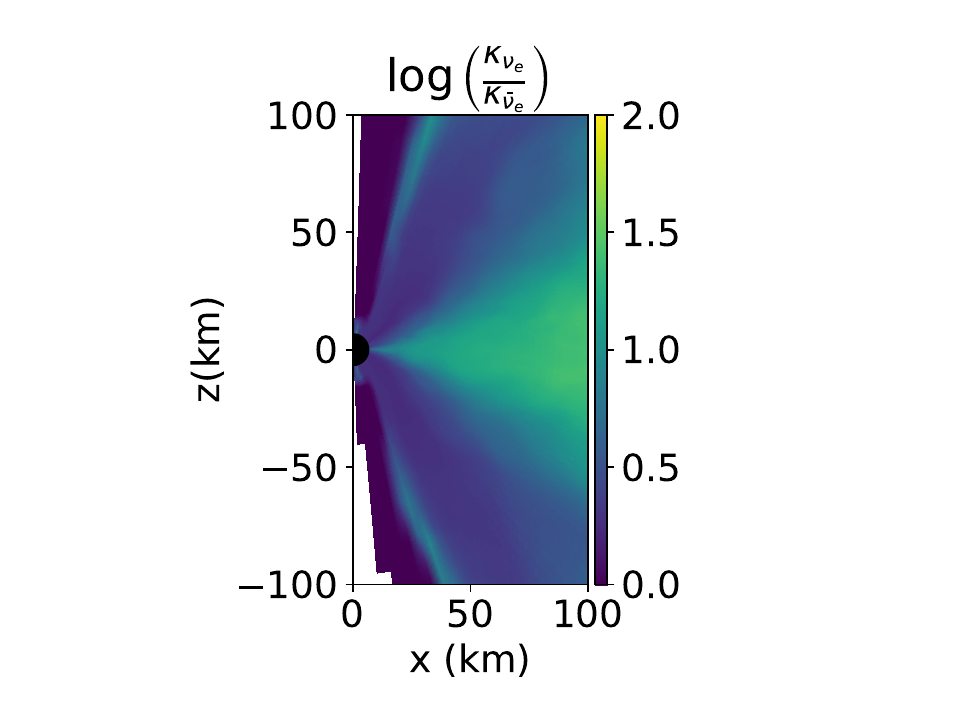}
    \caption{Top panel: 11ms snapshot. At these early times, $\kappa_{\nu_e}$ are many orders of magnitude greater than $\kappa_{\bar{\nu}_e}$ in a funnel shaped region around the disk. Bottom panel: $\sim 65$ ms snapshot. $\kappa_{\nu_e}$ and $\kappa_{\bar{\nu}_e}$ become more similar, but $\kappa_{\nu_e}$ is still about 1--2 orders of magnitude greater than $\kappa_{\bar{\nu}_e}$ near the disk.} 
    \label{fig7}
\end{figure*}

We now turn to a discussion of the structure and evolution of ELN-XLN crossings.
Figs. \ref{fig6} and \ref{fig7} show the emissivities ($\epsilon$) and opacities ($\kappa$) of $\nu_e$ and $\bar{\nu}_e$'s for 11 ms (top row) and 65 ms (bottom row). The first two plots on each row of Fig. \ref{fig6} shows $\log{\epsilon_{\nu_e}}$ and $\log{\epsilon_{\bar{\nu}_e}}$ and the third plot shows the ratio $\log{\left(\frac{\epsilon_{\nu_e}}{\epsilon_{\bar{\nu}_e}}\right)}$. At 11ms, $\epsilon_{\nu_e}$ can get up to about 4 orders of magnitude greater than $\epsilon_{\bar{\nu}_e}$ around the equator. $\epsilon_{\nu_e}$ clearly dominates over $\epsilon_{\bar{\nu}_e}$ in a funnel shaped region (shaded in blue). For consistency of notation, we will call this blue region the `plume'. The white band surrounding the plume is where $\nu_e$ and $\bar{\nu}_e$ emissivities have the same value. Beyond this conical shaped plume, the overall emissivities decrease in magnitude and $\bar{\nu}_e$ emission dominates (red shaded region). Since the absolute emissivities are dominated by this plume, the neutrinos that contribute to the crossings also originate from within this plume as we will discuss shortly. For $t\sim 65$ ms, $\epsilon_{\nu_e}$ still dominates $\epsilon_{\bar{\nu}_e}$ around the equator as can be seen clearly in the bottom right plot of Fig. \ref{fig6}. The ratio $\frac{\epsilon_{\nu_e}}{\epsilon_{\bar{\nu}_e}}$, in the plume is much smaller however, with $\epsilon_{\nu_e}$ only exceeding $\epsilon_{\bar{\nu}_e}$ by at most 1 order of magnitude. The shape of the plume is also different than for the 11ms case since it is more spread out. We will discuss shortly that this emissivity structure, combined with the opacity structure in Fig. \ref{fig7} is crucial for setting the structure and evolution of the ELN-XLN crossings.

Fig. \ref{fig7} highlights the drastic difference between the opacities $\kappa_{\nu_e}$ and $\kappa_{\bar{\nu_e}}$ for the 11 ms case. $\kappa_{\nu_e}$ is many orders of magnitude greater than $\kappa_{\bar{\nu}_e}$ inside the plume region. Combining the 11 ms emissivities in Fig. \ref{fig6} and opacities in Fig. \ref{fig7}, we see that within the plume, $\nu_e$ emission is dominant, but $\nu_e$s also have to traverse a much more optically thick region to reach anywhere in the simulation volume. The $\bar{\nu}_e$'s originating from within the plume on the other hand, will traverse less optical depth even though they are produced in subdominant numbers inside the plume. For the 65 ms case, $\kappa_{\nu_e}$ still dominates over $\kappa_{\bar{\nu}_e}$, but the difference in opacities is decreased. 

\begin{figure*}
    \centering
     \includegraphics[width=\columnwidth]{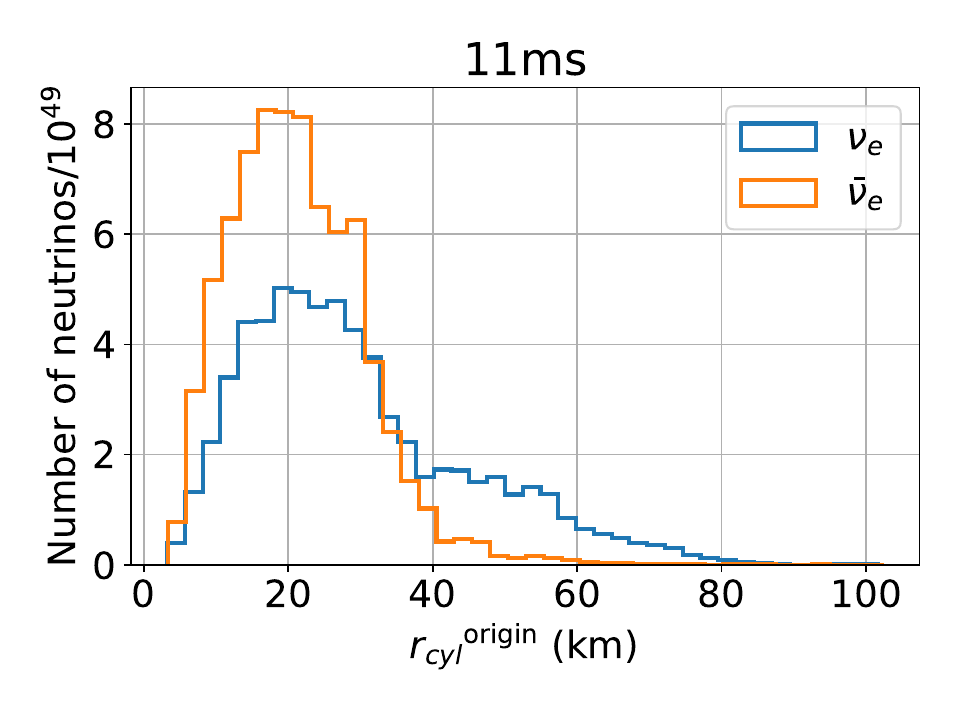}
     \includegraphics[width=\columnwidth]{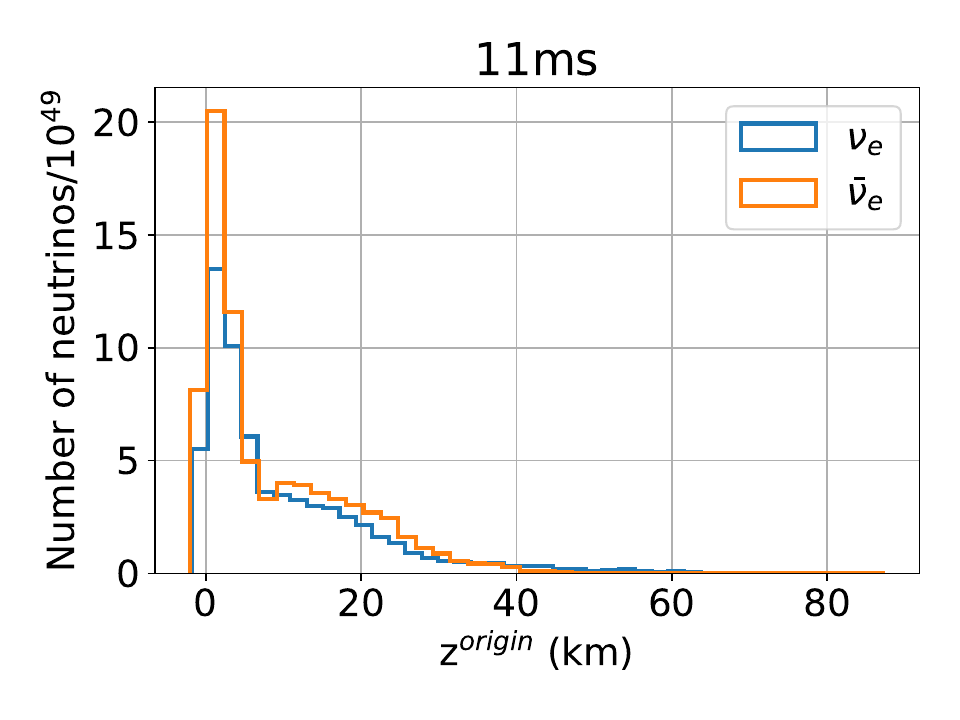}
     
\includegraphics[width=\columnwidth]{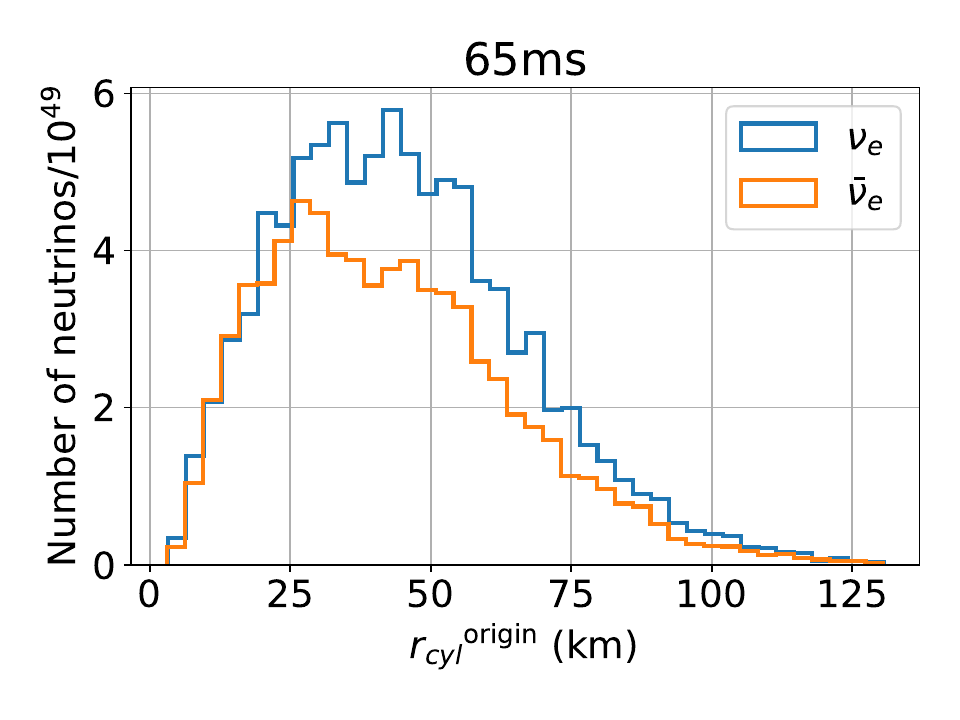} 
\includegraphics[width=\columnwidth]{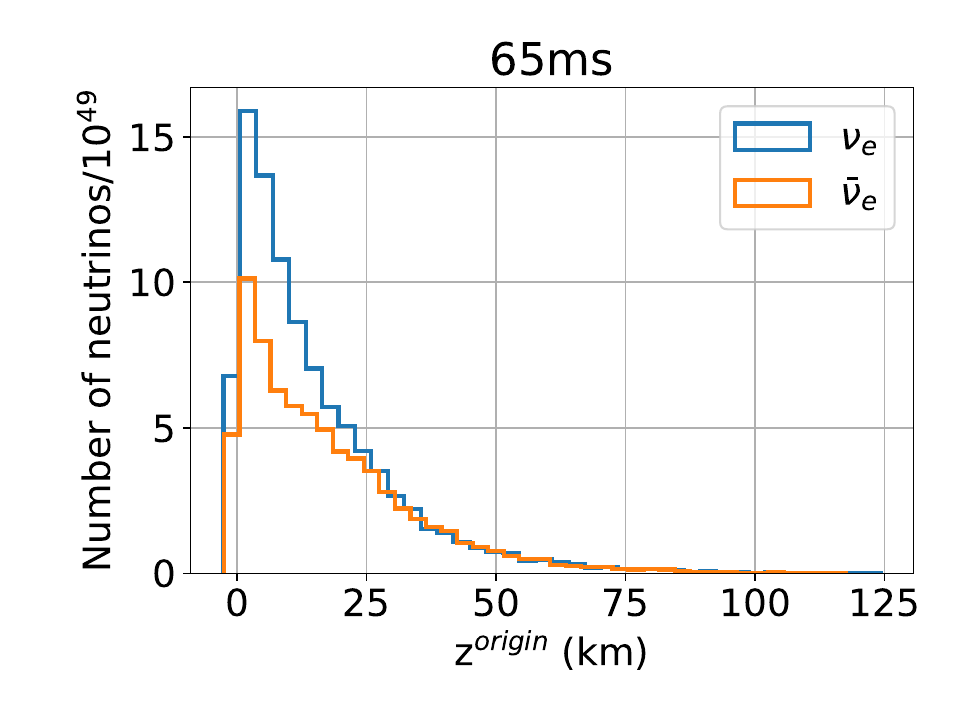}
    \caption{Origin locations ($r^{cyl}, z$) of $\nu_e$s and $\bar{\nu}_e$s that are present in the $r=60$ km, $\theta^{polar} = 45 \degree$ control volume at 11 and 65 ms. For both of the time snapshots, most of the emission happens very close to the disk as can be seen from the $z^{origin}$.} 
    \label{fig8}
\end{figure*}

To demonstrate how emissivities and opacities impact the formation of crossings, we consider a control volume located at $r =$ 60 km and $\theta = 45\degree$. An ELN-XLN crossing does not exist for this control volume for the 65 ms case (bottom left panel of Fig. \ref{fig:65msangular}), while a crossing exists for the 11 ms case (right panel of Fig. \ref{fig:11msangular}). Fig. \ref{fig8} shows the origin locations ($\mathrm{r}_{cyl}^{\mathrm{origin}}, \mathrm{z}^{\mathrm{origin}}$) of $\nu_e$'s and $\bar{\nu}_e$'s present inside the control volume for both 11 ms (top panels) and 65 ms (bottom panels). For both time points, it is clear that $\nu_e$s and $\bar{\nu}_e$s reaching the control volume are predominantly emitted from the central regions ($\mathrm z^{\mathrm{origin}} \leq$  10 km). At 11 ms, more $\bar{\nu}_e$s from within $r_{cyl} \leq 30$ km make it to the control volume than $\nu_e$s. This is 
consistent with 
the greatly reduced opacities experienced by $\bar{\nu}_e$s compared to $\nu_e$s on their way out to the control volume (top panels of Fig. \ref{fig7}). However, since $\epsilon_{\nu_e} > \epsilon_{\bar{\nu}_e}$ in the relevant ranges of $r_{cyl} ~\mathrm{and} ~z$ (top panel of Fig. \ref{fig6}), $\nu_e$s still arrive in greater numbers from $r_{cyl} > 40$ km, as a combination of higher emissivities and shorter optical depth due to decreased path length. As a consequence, at 11ms, one has more $\bar{\nu}_e$s contributing at forward angles but a tail with higher $\nu_e$ fraction beginning as one moves slightly away from the most forward angles. An intricate interplay of opacities and emissivities therefore leads to the ELN-XLN crossing. In contrast to this, for $t\sim 65$ ms, $\nu_e$s overwhelm $\bar{\nu}_e$s (bottom panel of Fig. \ref{fig8}) in the control volume as a consequence of reduced opacities (Fig. \ref{fig7} bottom panel).

\begin{figure}
    \centering
    \includegraphics[width=\columnwidth]{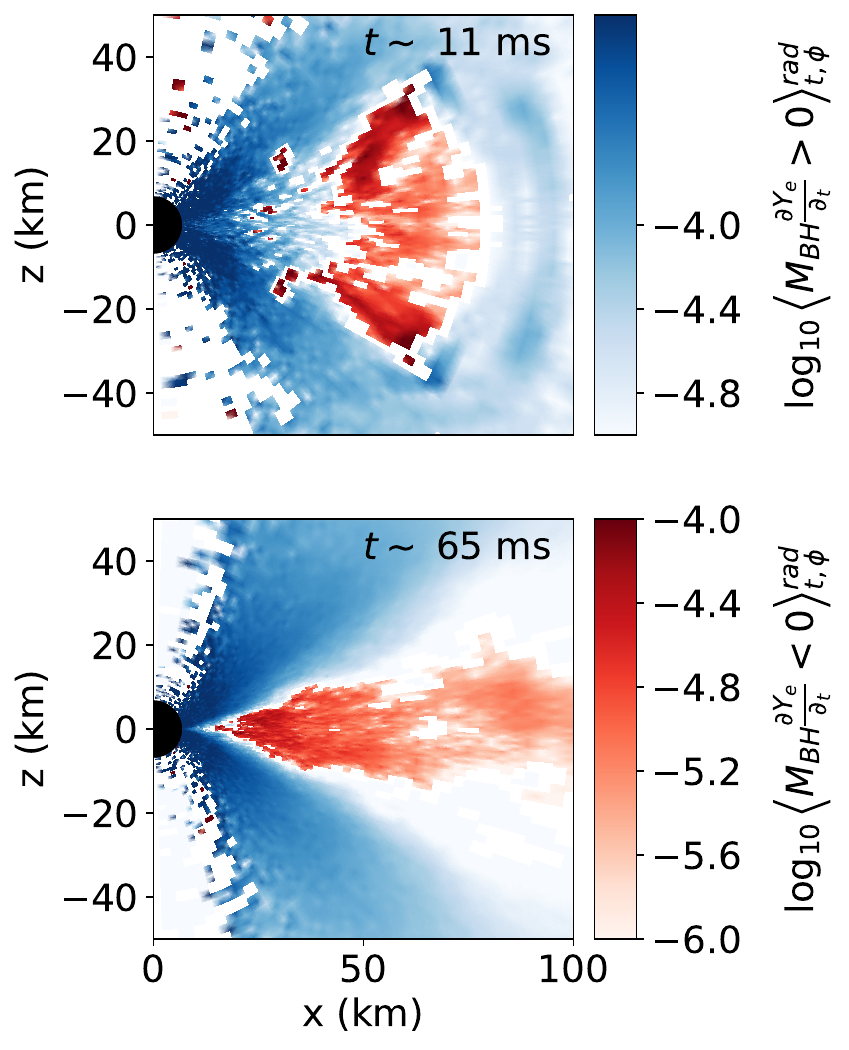}
    \caption{Lagrangian derivative of $Y_e$ due to emission or absorption of neutrinos for 11 ms (top) and 65 ms (bottom). Blue regions correspond to a decrease in $Y_e$ and red for a decrease. Averaged over the azimuthal angle $\phi$ and in time from 0 to 20 ms (top) and 50 to 70 ms (bottom). }
    \label{fig9}
\end{figure}

A natural question arises here. For the 65 ms case, $\nu_e$s overwhelm $\bar{\nu}_e$s in the control volume (bottom panel of Fig. \ref{fig8}). Similarly, for the 11 ms case, $\nu_e$s overwhelm $\bar{\nu}_e$s for $r_{cyl} > 40$ km (top panel of Fig. \ref{fig8}). This indicates that there are regions in the disk which are deleptonizing. However, shouldn't the expanding NS merger disk be leptonizing? To investigate this, we plot the Lagrangian derivative of the electron fraction due to emission or absorption of neutrinos in Fig. \ref{fig9} (blue for an increase in $Y_e$ and red for a decrease). The results are averaged over azimuthal angle $\phi$ and $t$ from 0 to 20 ms (top, representative of roughly $t\sim11$ ms) and from 50 ms to 70 ms (bottom, representative of roughly $t\sim 65$ ms). The time averaging is done to minimize effects of fluid turbulence. The 11 ms snapshot clearly shows that even though the disk as a whole is leptonizing (blue), but there is a region between $x\sim 40-80$ km which is deleptonizing and similarly for the 65 ms case. This deleptonization is coming from material traveling into the inner regions of the disk, which are hotter and denser.

Robust crossings which were seen all across the disk in Fig. \ref{fig:65msequator} for the 65 ms snapshot can now be successfully explained. $\kappa_{\nu_e} \gg \kappa_{\bar{\nu}_e}$ results in a higher fraction of radially directed $\bar{\nu}_e$s which forms the peak of the ELN-XLN crossing while the higher fraction of $\nu_e$s in the distribution tail is contributed by local $\nu_e$s with $\epsilon_{\nu_e} >> \epsilon_{\bar{\nu}_e}$. 
For all results presented in this section, we have verified that increasing the resolution of the control volume (making them smaller in size) does not change any of the above conclusions, albeit the statistical confidence of the results decreases.

\section{Discussion}

Broadly speaking, the electron fraction in outflows from simulations of post-merger disks depends on the balance of several factors. Typically more sophisticated transport schemes raise the electron fraction, especially at early times, when re-leptonization due to absorption is significant. On the other hand, longer, high-fidelity simulations, such as \cite{Sprouse_Emergent_2024} suggest a slow, low-$Y_e$ massive outflow eventually emerges. Fast flavor oscillations also typically lower $Y_e$, as electron neutrinos will be preferentially emitted, and then oscillate into another flavor before they can be re-absorbed \cite{li_NeutrinoFastFlavor_2021,just_FastNeutrinoConversion_2022}.

The electron fraction impacts the nucleosynthetic yields in part by modifying the number of free neutrons available for capture and thus the the contribution of mergers to the heavy element abundances in the universe.  The nucleosynthesic outcome also determines the electromagnetic counterpart from a merger. Even if the total lanthanide abundance of the outflow is large, if there is a fast low-lanthanide component to the outflow, the kilonova will have a blue component. The cut-off between lanthanide production and no-lanthanide production is in the region of $Y_e \sim 0.2 - 0.25$ depending on the timescale and entropy of the outflow and a fair amount of the ejected material winds up in this range, see Fig. \ref{fig:densityandye} for the $Y_e$ in the remnant.  Small differences in the angular distribution of neutrinos will result in differences in predicted $Y_e$ values which will propagate through to differences in the prediction of r-process observables.

With this in mind, we compare our work with Ref. \cite{just_FastNeutrinoConversion_2022} who search for ELN-XLN crossings by postprocessing a black hole accretion disk simulation with an M1 moment-based neutrino transport. The $Y_e$ distribution of this simulation is consistent with our simulation \cite{miller_FullTransportModel_2019}. They report ubiquitous crossings at early time, but they find a somewhat greater extent of ELN-XLN crossings at later times (Fig 1 of \cite{just_FastNeutrinoConversion_2022}), as compared with  our finding that the regions hosting ELN-XLN crossings shrink with time. We attribute this difference to the fact that traditional two-moment approaches cannot accurately capture the difference in the curvatures of the $\nu_e$ and $\bar{\nu}_e$ angular distributions presented in Figs. \ref{fig:65msequator}, \ref{fig:65msangular} and \ref{fig:11msangular} because it only takes two moments and an imposed closure relationship into account, while the Monte Carlo distribution can accurately assess the depth of the crossing as a consequence of having full angular information. This suggests that having full access to the neutrino distribution may be impactful in the prediction of observables. 

However, we note that since we are using a postprocessing framework we cannot make a definitive determination about how impactful these full distributions will eventually prove to be for element synthesis.   In particular, the fast flavor transformation will alter the neutrino angular distributions so as to render the neutrino flavor field no longer flavor unstable, e.g. \cite{Richers:2021xtf}. This has several implications.  One implication is that some regions where we predict crossings will not have crossings in a fully consistent calculation, as both neutrino and antineutrino distributions depend heavily on neutrinos coming from the regions of the disk that have already experienced the fast flavor conversion.  A second implication is that crossings may be present in the regions where we predict no crossings.  Our prediction of no fast flavor conversion relied on an excessive of neutrinos in all angular bins, but flavor conversion near the center of the disk, where many of these neutrinos originate, may reduce the number of neutrinos that arrive in our no-crossing regions.   These caveats provide additional motivation for developing techniques that can be used in dynamical simulations that perform both Monte Carlo neutrino transport and flavor conversion simultaneously, e.g.  \cite{Nagakura:2023jfi}.

\begin{figure}[t]
    \centering
    \includegraphics[width=\columnwidth]{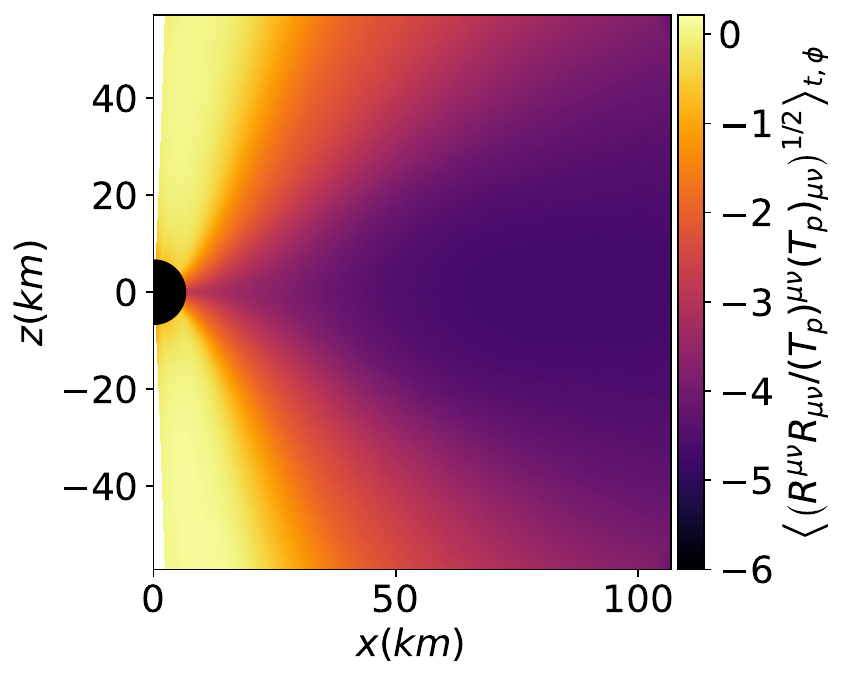}
    \caption{Ratio of the norm of the radiation stress-energy tensor $^\mu_{\ \nu}$ to the norm of the plasma stress energy tensor $(T_p)^{\mu}_{\ \nu}$. The tensors are extracted from a flow field averaged over azimuthal angle $\phi$ and times greater than 25 ms.}
    \label{fig:stress:ratios}
\end{figure}

While fast flavor transformations are very likely to impact nucleosynthesis in the disk outflow, we anticipate that they are unlikely to impact the dynamics. The reason for this is that in the absence of a neutron star remnant, the optical depths in post-merger disk are still relatively low (of order 1) in the region where significant net momentum deposition might occur, and the impact of absorption of neutrinos on the four-momentum of the gas is relatively low, of order a few percent. Figure \ref{fig:stress:ratios} compares the norm of the stress-energy tensor of the neutrino field to the stress-energy tensor of the gas. Except in the jet region, the stress-energy tensor for radiation is of order a few percent of the stress-energy tensor of the gas, suggesting that neutrino radiation pressure and gradients thereof are subdominant in disk dynamics. 

\begin{figure}[t]
    \centering
    \includegraphics[width=\columnwidth]{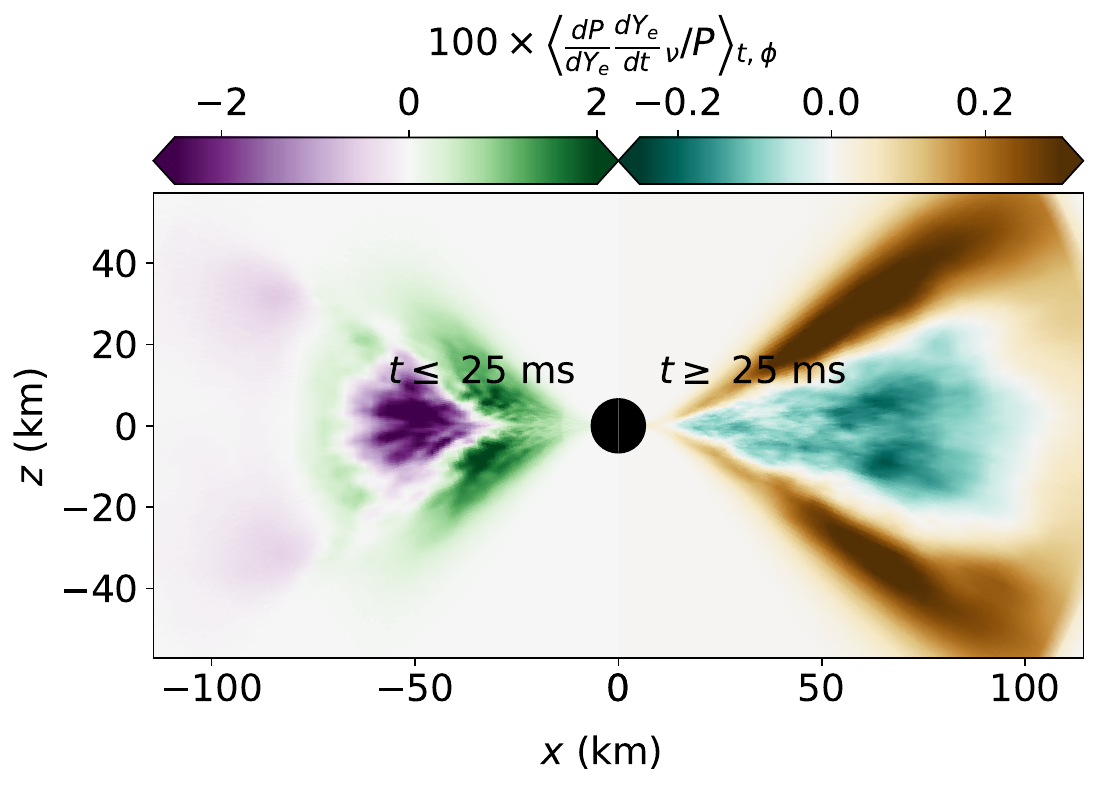}
    \caption{The Lagrangian derivative in pressure due to change in electron fraction from the emission and absorption of neutrinos. Averaged over time and azimuthal angle $\phi$. Left: averaged for times less than 25 ms. Right: averaged for times greater than 25 ms. Note the difference in scales.}
    \label{fig:dPdYe}
\end{figure}

Also along these lines, we need to consider the relative scale of the pressure gradient introduced by the change in $Y_e$ due to emission and absorption since these pressure gradients are indirectly affected by neutrino flavor transformation.  Figure \ref{fig:dPdYe} plots the Lagrangian derivative in pressure due to the change in electron fraction due to this effect and we can see that this effect is of a similar scale. At early times (left panel), of order a few percent of the pressure is set by gain or loss of electron number due to the radiation field. At late times (right panel), this effect becomes negligible.

\section{Conclusions}

In this paper, we have analyzed a black hole accretion disk where the equations of general relativistic ideal magnetohydrodynamics have been solved using \texttt{$\nu$blight} code. An explicit Monte Carlo scheme has been used to evolve the neutrinos taking into account emission, absorption and scattering with matter. The radiation field is generated self-consistently with the simulation but it does not include the feedback from the neutrino oscillations. We post-processed the simulation results and focused our attention on finding electron lepton number crossings in this system as the disk evolves in time. The motivation for this study is the crucial fact that having ELN-XLN crossings is a reliable indicator of the existence of fast flavor instabilities, which, if present can dramatically impact the nucleosynthesis and ultimately the kilonova signatures of binary neutron star mergers. 

Within the postprocessing framework, we have access to the full neutrino angular distributions as well as sufficient statistics to determine the presence of crossings. 
An advantage of our approach over more traditional moment-based approaches is that in moment based approaches the full angular distribution must be gleaned from two moments and an imposed closure relationship. We find that statistically significant ELN-XLN crossings are present in the system at both early and late times. This is consistent with previous reports in the literature \cite{wu_FastNeutrinoConversions_2017,wu_ImprintsNeutrinopairFlavor_2017a,li_NeutrinoFastFlavor_2021,just_FastNeutrinoConversion_2022}.  However, we also find the geometric structure of the crossings evolves with time, and at later times the region where crossings are present shrinks to occur only close to the equator. This is a new finding that we attribute to the additional angular information that is retained in the Monte-Carlo neutrino transport method. We have provided a detailed analysis of the radiation field, disentangling the emission and absorption effects that are giving rise to this time-evolving ELN-XLN crossing structure. 

Some of the distributions we present would be difficult to reproduce with a classical moment method, suggesting that having access to the full neutrino angular distribution can be important for understanding the true nature of fast flavor oscillations in black hole accretion disks, and ultimately the heavy element nucleosynthesis in binary merger systems.  The ultimate goal is to put flavor transformation together with neutrino transport using theoretically motivated equations.  An approach with a clear separation of scales stems neutrino quantum kinetics \cite{Volpe:2013uxl,Vlasenko:2013fja,Richers:2019grc}, although the potential for effects that may neglected in this approach are currently under study \cite{Patwardhan:2022mxg}.  Various computational methods are being developed to do this, e.g.  \cite{Richers:2022bkd,Grohs:2022fyq,Grohs:2023pgq,Johns:2024dbe}, and such methods are beginning to be used in explosive systems \cite{Xiong:2022vsy,Xiong:2024pue}. Our work serves as a motivation to extend such efforts to incorporate the feedback effects of fast flavor oscillations in a Monte Carlo simulations.

\section{Acknowledgements}
We thank Sherwood Richers, Luke Johns and Daniel Kasen for their insightful comments. The work of PM is supported in
part by the Neutrino Theory Network Program Grant
under award number DE-AC02-07CHI11359. JMM's contribution was supported through the Laboratory Directed Research and Development program under project number 20220564ECR at Los Alamos National Laboratory (LANL). This research also used resources provided by the LANL Institutional Computing Program. LANL is operated by Triad National Security, LLC, for the National Nuclear Security Administration of U.S. Department of Energy (Contract No. 89233218CNA000001). This work is approved for unlimited release with report number LA-UR-24-23662.  We acknowledge support from the NSF (N3AS PFC) grant No. PHY-2020275, as well as from U.S. DOE contract Nos. DE-FG0202ER41216 and DE-SC00268442 (ENAF). 
This work was partially supported by the Office of Defense Nuclear Nonproliferation Research \& Development (DNN R\&D), National Nuclear Security Administration, U.S. Department of Energy. 
This work is performed in part under the auspices of the U.S. Department of Energy by Lawrence Livermore National Laboratory with support from LDRD project 24-ERD-02.

\clearpage
\bibliography{FFI_bib.bib}

\end{document}